\documentclass[12pt]{nature}

\usepackage{graphicx}
\usepackage{times}
\usepackage{xcolor}
\usepackage{multibib}

\newcites{Main}{References}
\newcites{Method}{Methods References}

\title{Title: Resolving acceleration to very high energies along the Jet of Centaurus A} 

\author{
Author: The H.E.S.S. Collaboration\\
\footnotesize{Correspondence to: contact.hess@hess-experiment.eu}
\\
\footnotesize{The full author list with affiliations can be found at the 
end of this paper}
}

\date{}

\def\deg{\hbox{$^\circ$}}

\begin{document} 

\baselineskip24pt

\maketitle 

%

\noindent
\textbf{Summary:\\ 
The nearby radio galaxy Centaurus A belongs to a class of Active Galaxies that are very luminous at radio 
wavelengths. The majority of these galaxies show collimated relativistic outflows known as jets, that extend 
over hundreds of thousands of parsecs for the most powerful sources. Accretion of matter onto the central 
super-massive black hole is believed to fuel these jets and power their emission \citeMain{2019ARA&A..57..467B}, 
with the radio emission being related to the synchrotron radiation of relativistic electrons in magnetic fields.
The origin of the extended X-ray emission seen in the kiloparsec-scale jets from these sources is still a matter 
of debate, although Centaurus A's X-ray emission has been suggested to originate in electron synchrotron processes 
\citeMain{1981ApJ...251...31F,2002ApJ...569...54K,2019ApJ...871..248S}. The other possible explanation is inverse 
Compton scattering with CMB soft photons \citeMain{2001MNRAS.321L...1C,2006ARA&A..44..463H,2016ApJ...816L..15S}. 
Synchrotron radiation needs ultra-relativistic electrons ($\sim50$ TeV), and given their short cooling times, 
requires some continuous re-acceleration mechanism to be active \citeMain{2017ApJ...842...39L}. 
Inverse Compton scattering, on the other hand, does not require very energetic electrons, but requires jets that 
stay highly relativistic on large scales ($\geq$1 Mpc) and that remain well-aligned with the line of sight. Some 
recent evidence disfavours inverse Compton-CMB models \citeMain{2016Galax...4...65G,2017ApJ...849...95B,
2018A&A...612A.106S,2018ApJ...856...66M}, although other evidence seems to be compatible 
with them \citeMain{2017MNRAS.466.4299L,2019ApJ...883L...2M}. In principle, the detection of extended gamma-ray 
emission, directly probing the presence of ultra-relativistic electrons, could distinguish between these options, 
but instruments have hitherto been unable to resolve the relevant structures. 
At GeV energies there is also an unusual spectral hardening in Centaurus A \citeMain{2013ApJ...770L...6S,2018_CenACore}, 
whose explanation is unclear. Here we report observations of Centaurus A at TeV energies that resolve its large-scale 
jet. We interpret the data as evidence for the acceleration of ultra-relativistic electrons in the jet, and favour 
the synchrotron explanation for the X-rays.}\\


Centaurus A is the closest known radio galaxy at a distance of $3.8\,$Mpc \citeMain{2010PASA...27..457H}, offering a unique 
opportunity to better resolve the processes at play in the jets. Its radio morphology exhibits a one-sided, kpc-scale radio jet 
extending out from the nucleus at a position angle of $55^{\circ}\pm 7^{\circ}$ (measured counter-clockwise from North in the 
equatorial system), two inflated inner radio lobes extending to about $5\,\mathrm{kpc}$ north and south of the nucleus, and 
extended low surface-brightness structures ("giant lobes") with a size of a few hundred kpcs \citeMain{1983ApJ...273..128B,
1998A&ARv...8..237I}. Along much of the length of the radio jet, from within $50\,\mathrm{pc}$ to $4\,\mathrm{kpc}$, X-ray 
emission at $\sim10^3\,$eV has been detected with the Chandra satellite \citeMain{2000ApJ...531L...9K,2002ApJ...569...54K}. 
The source is also positionally close to an ultra-high energy cosmic-ray hotspot \citeMain{2018ApJ...853L..29A}. 


Gamma-rays from the central part of Centaurus A were first detected in the MeV-GeV range by the Compton Gamma-Ray 
Observatory (CGRO) \citeMain{1999ApJS..123...79H} 
and above hundreds of GeV by the High Energy Stereoscopic System (H.E.S.S.) \citeMain{2009ApJ...695L..40A}. 
The kpc-scale jet, however, was so far not resolved. Much of the observed emission has previously 
been associated with jet regions close to the black hole \citeMain{2001MNRAS.324L..33C,2008A&A...478..111L}. 
On larger scales, the giant lobes of Centaurus A have been seen in the GeV range by the \textit{Fermi} Large Area Telescope (LAT) \citeMain{2010Sci...328..725A}, 
making this source the first extended object in the extragalactic GeV sky.

To probe a possible extension at very high energy (VHE, $E >100\,$GeV) gamma-ray energies, we selected a total of 202 hours of H.E.S.S \citeMain{HessCrab} observations 
of Centaurus A between 2004 and 2016, corresponding to high-quality data suitable for extension measurements (see Methods~I 
for more information about the analysis). 
The analysis configuration applied to the reconstructed data corresponds to a compromise between achieving a good angular resolution whilst 
retaining a sufficiently low energy threshold and good gamma-ray sensitivity for a faint source like Centaurus A. The detection 
significance is $13.1\sigma$ (compared to $12\sigma$ with 213 hours of observation time as found in \citeMain{2018_CenACore}) 
at energies above $240\,$GeV. 
The Point Spread Function (PSF) for the given dataset and configuration was simulated within the framework described in \citeMain{2017_RunWise}. 
It was convolved with different source models and fitted to the data using \textit{Sherpa} \citeMain{2001_Sherpa}. 
We compare the different best-fit models by using a test statistic (TS) value (see Methods~I) 
as figure of merit. In addition to the assumption of point-like emission, we fit the data with a radially symmetric Gaussian as well as an 
elliptical Gaussian model. Compared to a point-like source, the radially symmetric Gaussian is preferred with a TS of $6.1$, and 
the elliptical Gaussian with $\mathrm{TS} = 19.4$. The considerable difference of the TS values implies a strong preference for 
the elliptical over the radially symmetric model.

\begin{table}
\centering
\begin{tabular}{lcc}
  \hline
  \hline
  \noalign{\smallskip}
  Parameter &  Value & Statistical Error \\
  \hline
  \noalign{\smallskip}
  $\sigma_{\mathrm{maj}}$ ($^{\circ}$) & $0.041$ & $0.006$ \\ 
  \noalign{\smallskip}
  $\sigma_{\mathrm{min}}$ ($^{\circ}$) & $< 0.013$ & - \\ 
  \noalign{\smallskip}
  Ellipticity $\varepsilon$ & $0.92$ & $+\, 0.08 - 0.23$ \\ 
  \noalign{\smallskip}
  Position Angle $\varphi$ ($^{\circ}$) & $43.4$ & $+\,7.7 - 7.2$ \\
  \noalign{\smallskip}
  \hline
\end{tabular}
\caption{\textbf{Best-fit parameters of the elliptical Gaussian model.} The width of the semi-major axis $\sigma_{\mathrm{maj}}$ has a position angle $\varphi$, 
measured counter-clockwise from North.} 
\label{Tab_FitStats}
\end{table}

The best-fit parameters of the elliptical Gaussian are given in Table~\ref{Tab_FitStats}. The position angle of the semi-major axis is compatible with that of the 
radio and X-ray jets \citeMain{1983ApJ...273..128B,2003ApJ...593..169H}. 

\begin{figure}
\centering
\includegraphics[width=0.7 \textwidth]{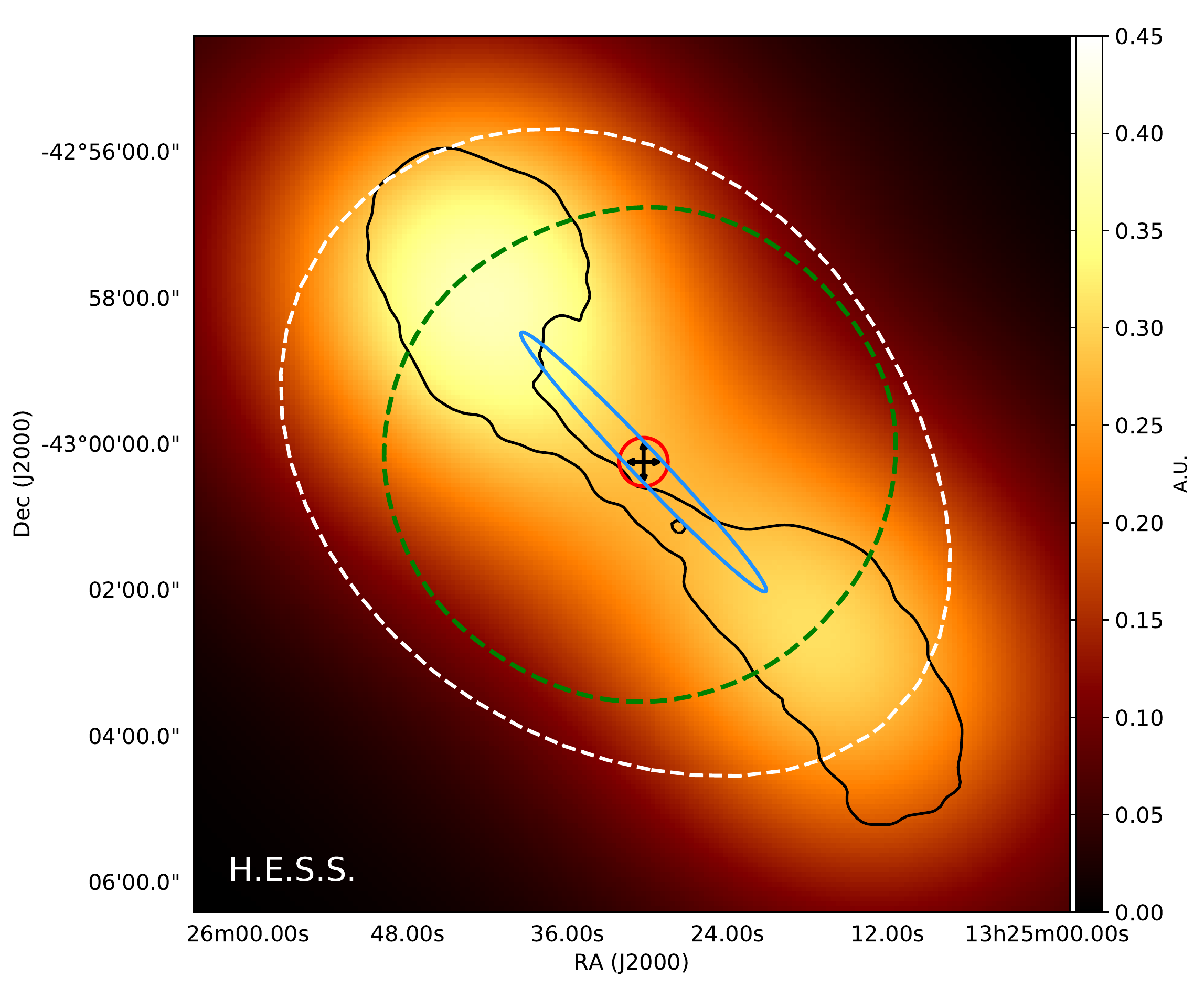}
\caption{\textbf{Multiwavelength image of Centaurus A.} The color map represents the radio ($21\,$cm) VLA map of Centaurus A \protect\citeMain{1996_Condon_VLA}, 
after convolution with the H.E.S.S. PSF and an additional oversampling with a radius of $0.05^{\mathrm{\circ}}$. 
Contours of the unconvolved VLA map, with levels adjusted to highlight the core (corresponding to $4\,\mathrm{Jy}/\mathrm{beam}$) as well as the kpc-scale jet ($0.5\,\mathrm{Jy}/\mathrm{beam}$), are drawn in black. 
The VHE gamma-ray morphology of Centaurus A is represented by a white dashed contour which is derived from the $5\sigma$ excess significance level of the H.E.S.S. sky map, also after oversampling with a radius of $0.05^{\mathrm{\circ}}$. 
The result of the best fit of an elliptical Gaussian to the H.E.S.S. measurement is shown in blue by its $1\sigma$ contour which corresponds to a model containment fraction of $39\%$. The $1\sigma$ statistical uncertainties of the fitted position are drawn as black arrows, and the estimated pointing uncertainties with a red circle. The dashed green line denotes the $68\%$ containment contour of the H.E.S.S. PSF.}\label{Map}
\end{figure}

This is further illustrated in Figure~\ref{Map}. The Gaussian width of the semi-major axis $\sigma_{\mathrm{maj}}$, together with 
the obtained ellipticity $\varepsilon = 1 - \sigma_{\mathrm{min}}/\sigma_{\mathrm{maj}}$ denotes the $39\%$ containment of 
measured gamma-rays from Centaurus A. The position of the best-fit model using J2000 coordinates is $\alpha = 
13\,\mathrm{h}\,25\,\mathrm{m}\,30.3\,\mathrm{s} \pm (1.4\,\mathrm{s})_{\mathrm{stat}} \pm (1.8\,\mathrm{s})_{\mathrm{sys}}$, 
$\delta =  -43\deg00'15'' \pm (15'')_\mathrm{stat} \pm (20'')_{\mathrm{sys}}$ (systematic pointing errors taken from \citeMain{2004_Gillesen}). 
This corresponds to a slight, insignificant offset of approximately $60''$ north-east from the position of the galaxy core \citeMain{1998AJ....116..516M}. 

\begin{figure}
\begin{center}
\includegraphics[width=0.8 \textwidth]{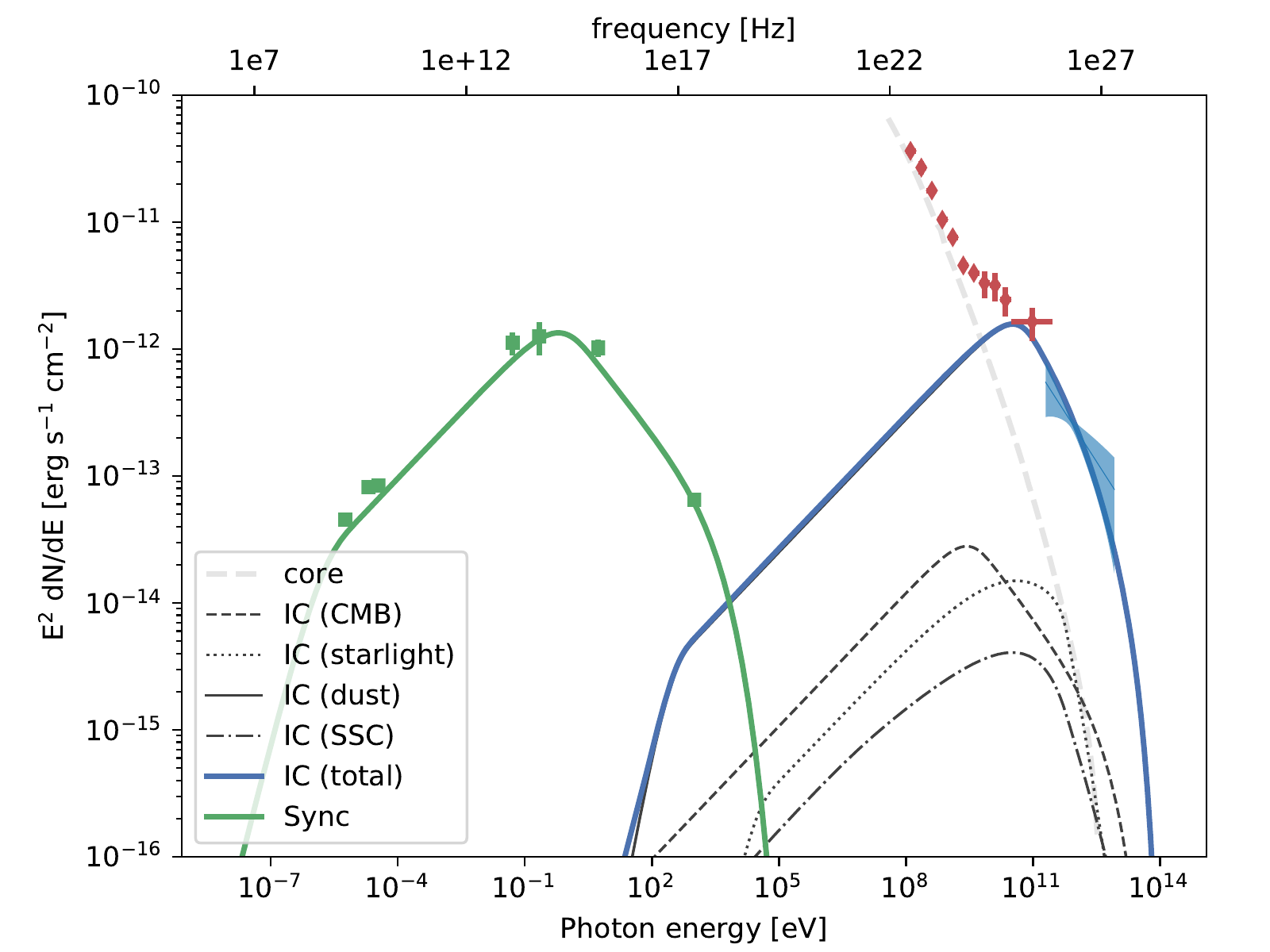}
\caption{\textbf{Spectral energy distribution of Centaurus A.} Observed and modelled spectral energy distribution (SED) from radio to 
gamma-ray energies for the inner, kpc-scale jet of Centaurus~A. The VHE emission is dominated by relativistic electrons with Lorentz factor $\gamma
 \geq 10^7$ IC upscattering dust photons to high energies (solid/blue curve). This emission from the kpc-scale jet makes a major contribution 
to the unexpected spectral hardening above a few GeV as seen by {\it Fermi}-LAT (red points) \protect\citeMain{2018_CenACore}. 
The lower-energy part of the gamma-ray spectrum (red points) is attributed to emission from the core (dashed line referring to a core model fit from \protect\citeMain{2018_CenACore}).
The green curve designates the synchrotron emission of the inferred broken power-law electron distribution in a magnetic field of characteristic strength $B=23\,\mu$G. 
The blue butterfly corresponds to the H.E.S.S. spectra, while green data points mark radio, infrared and X-ray measurements and reported uncertainties from the 
inner region of the Centaurus A jet (see Methods~II). 
A breakdown is provided of the full inverse Compton contribution, from the scattering of: Cosmic Microwave Background (CMB), the low-energy 
synchrotron jet emission, infrared emission from dust, and the starlight emission of the host galaxy. Data are from \protect\citeMain{2018_CenACore} 
and \protect\citeMain{2006MNRAS.368L..15H}, see Methods~II 
for further details.}\label{SED}
\end{center}
\end{figure}

The physical extension of the semi-major axis of the best-fit elliptical Gaussian exceeds $2.2\,$kpc, 
implying that a major part of the VHE emission arises on large scales, far away from the black hole. 
The derived alignment with the jet direction and the known spectral characteristics are in line with models where the VHE 
emission originates from inverse Compton (IC) upscattering of 
low-energy photons by very energetic electrons accelerated along the jet \citeMain{2011MNRAS.415..133H,2019MNRAS.483.1003B,2019ApJ...878..139T}. 
Figure~\ref{SED} shows a reproduction of the spectral energy distribution (SED) from radio to gamma-ray energies for jet-scales close to 
$2.2\,$kpc (see Methods~II 
for details). 
The IC emission on these scales is dominated by upscattering of infrared photons emitted by dust, with the scattering occurring predominantly in the Thomson regime. 
Note that the considered large-scale model is not intended to reproduce the high-energy emission below a few GeV, as 
this part of the SED is usually attributed to emission from the core. 

Regardless of specific details, the observed VHE extension provides the first direct evidence for the presence of ultra-relativistic electrons 
with Lorentz factors $\gamma \sim(10^7-10^8)$ within an extragalactic large-scale jet (see Methods, Extended Data Figure~\ref{SED_compare} for details). 
Assuming a synchrotron origin, the inferred X-ray spectral slope translates into a photon index $\simeq2.4$ which is close to that derived from Chandra observations (2.29 $\pm$ 0.05 and 2.44 $\pm$ 0.07 for the inner and middle region, respectively) \citeMain{2006MNRAS.368L..15H}. 
The results thus substantiate the synchrotron interpretation of the X-ray emission seen in the large-scale jet of Centaurus A, which was 
originally motivated largely by similarities between the radio and X-ray morphologies \citeMain{1981ApJ...251...31F,1983ApJ...273..128B}.  
Given that the synchrotron lifetimes of these extremely energetic electrons can be as low as a few hundred years, i.e. considerably less 
than the travel time down the jet which is of order of thousands of years, the detection of extended X-ray emission on kpc scales related 
to synchrotron emission requires the operation of an efficient, extended or distributed (re)acceleration mechanism far away from the black hole, 
such as stochastic or shear particle acceleration \citeMain{2017ApJ...842...39L} (see Methods~II). 

Interestingly, IC emission of the kpc-scale jet could make a major contribution to the unexpected spectral hardening \citeMain{2013ApJ...770L...6S,
2018_CenACore} seen in the high-energy gamma-ray emission of Centaurus A (see Figure~\ref{SED}).  
With its superior resolution and sensitivity, the Cherenkov Telescope Array (CTA) will in a few years time be able to probe deeper into the 
VHE extension and to search for potential VHE variability that would impose constraints on the ratio of the extended gamma-ray flux to 
the one from the core region \citeMain{2019_ScienceWithCTA}.

Our findings have important implications for understanding the diversity of gamma-ray emitters and the cosmic energy input at VHE energies. 
Observationally, the current measurement of the VHE extension in Centaurus A became possible only due to the unique 
proximity of the source and the presence of a significant amount of dust within the source (see Methods for more details). 
From a physical point of view however, the kpc-scale jet in Centaurus A is not very exceptional, be it with respect to its jet power, length, or speed. 
This suggests that ultra-relativistic electrons could be ubiquitously present in the large-scale jets of radio-loud active galaxies, 
making these jets a highly promising class of TeV emitters. 
To date, most extragalactic sources detected in VHE gamma-rays are of the BL Lac type \citeMain{2017AIPC.1792b0007T}. 
The gamma-ray emission in these objects is believed to originate on rather small ($\leq 1\,$pc) scales in fast parts of the jet viewed almost face-on. 
This is accompanied by a strong Doppler amplification of their emission, favouring their detection in the extragalactic sky. 
Quantitatively however, BL Lac objects make up only a small subclass of radio-loud active galaxies ($\leq3\%$), 
and their rest-frame (corrected for beaming) VHE contribution is usually relatively moderate. 
This in turn suggests that large-scale jets could provide a more relevant energy input into the intergalactic medium at TeV energies, 
even if they are in most cases not individually detectable at those energies \citeMain{2016Galax...4...65G}.
The results reported here demonstrate the unique power that VHE observations bring to the picture of Active Galactic Nuclei, allowing to 
probe multiple spectral components of local well-studied sources, which have previously evaded detection.

\bibliographystyleMain{Nature}
\bibliographyMain{CenA_Nature}

\section*{Acknowledgments}

The support of the Namibian authorities and of the University of Namibia in facilitating 
the construction and operation of H.E.S.S. is gratefully acknowledged, as is the support 
by the German Ministry for Education and Research (BMBF), the Max Planck Society, the 
German Research Foundation (DFG), the Helmholtz Association, the Alexander von Humboldt Foundation, 
the French Ministry of Higher Education, Research and Innovation, the Centre National de la 
Recherche Scientifique (CNRS/IN2P3 and CNRS/INSU), the Commissariat \`{a} l'\'{e}nergie atomique 
et aux \'{e}nergies alternatives (CEA), the U.K. Science and Technology Facilities Council (STFC), 
the Knut and Alice Wallenberg Foundation, the National Science Centre, Poland grant no. 2016/22/M/ST9/00382, 
the South African Department of Science and Technology and National Research Foundation, the 
University of Namibia, the National Commission on Research, Science \& Technology of Namibia (NCRST), 
the Austrian Federal Ministry of Education, Science and Research and the Austrian Science Fund (FWF), 
the Australian Research Council (ARC), the Japan Society for the Promotion of Science and by the 
University of Amsterdam. We appreciate the excellent work of the technical support staff in Berlin, 
Zeuthen, Heidelberg, Palaiseau, Paris, Saclay, T\"ubingen and in Namibia in the construction and 
operation of the equipment. This work benefited from services provided by the H.E.S.S. 
Virtual Organisation, supported by the national resource providers of the EGI Federation. 

\section*{Author Contribution Statement}
M.H., M.d.N., and D.S. analysed and interpreted the H.E.S.S. data and prepared the manuscript. F.R. and A.T. performed the modelling and prepared the manuscript. The entire H.E.S.S. collaboration contributed to the publication with involvement at various stages, from the design, construction and operation of the instrument to the development and maintenance of all software for data handling, data reduction and data analysis. All authors reviewed, discussed and commented on the present results and the manuscript.

\section*{Competing Interests}
The authors declare no competing interests.

\section*{Additional Information}
Correspondence and requests for materials should be addressed to contact.hess@hess-experiment.eu.
\\
Reprints and permissions information is available at www.nature.com/reprints.

\setcounter{figure}{0}
\setcounter{table}{0}
\makeatletter 
\renewcommand{\figurename}{Extended Data Figure}
\renewcommand{\tablename}{Extended Data Table}
\makeatother

\newpage
\pagenumbering{arabic}

\section*{Methods}
\subsection*{I. Data Analysis}
\label{som_methods}

The results presented in the main text were derived from data taken with the Imaging Atmospheric Cherenkov Telescopes (IACTs) of the 
High Energy Stereoscopic System (H.E.S.S.). This array is located in the Khomas Highland in Namibia at an altitude of $1800\,$m above sea level, 
providing excellent observing conditions for sources located in the southern hemisphere. 
It consists of four identical IACTs (CT1-4) each with an effective mirror area of $107\,\mathrm{m}^2$, in a square formation of side length $120\,$m, 
and a fifth, $614\,\mathrm{m}^2$ IACT (CT5) which was added to the centre of the array in 2012. 
From the location of H.E.S.S., Centaurus A is observable at zenith angles of at least $20\deg$. 
Observations are conducted in individual runs of up to 28 minutes duration.

In addition to the standard data quality selection \citeMethod{Method_HessCrab}, 
additional cuts were applied to ensure best quality and reliability of the simulated point spread function (PSF). 
Only runs with a \textit{Transparency Coefficient} (TC, see \citeMethod{Method_2014_TC}) in the range $0.8 < \mathrm{TC} < 1.1$ are used to guarantee good atmospheric conditions. 
To minimise the influence of inaccuracies of the pointing model, the maximum observation zenith angle was restricted to $\theta < 45\deg$. 
For the same reason, a comparison of the photomultiplier (PMT) currents with the expectation from bright stars in the respective field of view 
was used to sort out runs with deteriorated pointing accuracy. 
The final data set consists of 485 runs or $202\,$h of observations with an average zenith angle of $\theta_{\mathrm{avg}} = 23\deg$, 
taken between April 2004 and February 2016, the majority before August 2010. CT5 is not used in the present analysis.

The reconstruction and analysis was carried out with the \textit{Model} analysis technique \citeMethod{Method_2009_deNaurois}. 
In addition to the standard analysis configuration settings, only events with an estimated direction reconstruction uncertainty of less than $0.035\deg$ were considered, 
leading to an improved instrument PSF. With this configuration, 
Centaurus A is detected at a significance level of $13.1\sigma$ and a signal to background ratio of $0.49$, 
comparatively high for this source \citeMethod{Method_2018_CenACore}. 
The source flux is found to be constant within the measurement accuracy, showing no hint for variability.

Sky images of the measured gamma-like events (ON map) as well as the background (OFF map), 
the latter estimated using the ring background technique \citeMethod{Method_hess-background}, 
were generated in FITS format for the \textit{Sherpa} \citeMethod{Method_2001_Sherpa} morphology fit, using a finer binning of $0.005\deg \times 0.005\deg$. 
To obtain the PSF for the respective data set, run-wise simulations \citeMethod{Method_2017_RunWise} of a point-like gamma-ray source 
from the direction of the core of Centaurus A were conducted. The simulated data set was analysed with the same settings as the actual data. 
The assumed spectral index of Centaurus A is $\Gamma = 2.65$, and the total number of accepted simulated gamma-rays amounts to almost $800\, 000$ 
compared to roughly $500$ excess events in the data. 
Despite the fact that Centaurus A is generally observed towards the south with H.E.S.S. (implying a larger influence of the Earth magnetic field), 
the $68\%$ and $80\%$ PSF containment radii of this analysis are only $0.057\deg$ and $0.072\deg$, respectively. 
The corresponding sky map of reconstructed event directions, using the same binning of $0.005\deg \times 0.005\deg$, 
is used as PSF model in the morphological analysis. 
Since the effects that can cause a distortion of the PSF (such as, e.g., the Earth magnetic field or the array layout) 
are taken into account in the simulation framework, the map provides a reliable template which may be used to not only search for extension in general, 
but also for testing models without radial symmetry.

The \textit{Sherpa} fit was carried out using all bins within a radius of $0.5\deg$ around the position of the core of Centaurus A. 
For validation purposes, all fits were also done using a radius of $0.25\deg$ and yielded consistent results. 
Three nested models of increasing complexity were tested: The point-like hypothesis, a symmetric two-dimensional Gaussian, as well as an elliptical Gaussian model. 
More advanced or superimposed models were deliberately disregarded because of the limited statistics level. 
Each model was convolved with the PSF, added to the OFF map, and compared to the ON map. 
To account for potential background uncertainties, a constant offset was additionally allowed during the fit. 
The log-likelihood $\mathcal{L}$ was evaluated with the \textit{Cash} statistic \citeMethod{Method_1979_CashStats}, 
suitable for Poissonian event distributions, and minimized using the \textit{moncar} method \citeMethod{Method_1997_Moncar}, 
which is the best option to find optimum fit parameters in our case. 
Two models $a$ and $b$ are statistically compared via the test statistic value $\mathrm{TS} = \mathcal{L}_a - \mathcal{L}_b$. 
Statistical errors of a given model parameter were calculated by scanning around its minimum value, leaving all other parameters free, 
and searching for the value where $\mathrm{TS} = 1$. 
With an average background level of only around $0.7$ counts per bin, the overall bin-wise event statistics level is rather low, 
meaning that the $\mathrm{TS} = 1$ criterion corresponds to a conservative estimate of the $1\sigma$ confidence level.
The $TS = 1$ lower limit of this result converts to a physical extension of more than $2.2\,$kpc, 
providing a conservative estimate of the VHE extension of the jet.

As outlined in the main text, Centaurus A looks significantly extended to H.E.S.S., 
where an elliptical Gaussian model as illustrated in Figure~\ref{Map} is preferred by $\mathrm{TS} = 13.3$ over a radially symmetric Gaussian, 
and by $\mathrm{TS} = 19.4$ over a point-like assumption. 
These values notably relate to the chance coincidence of measuring an elliptical Gaussian along any direction, 
not taking into the account the alignment of our best-fit model with the one of the radio and X-ray jet. 
To check how the deviation from radial symmetry of the PSF affects the measured position angle 
$\varphi = (43.4 + 7.7_{\mathrm{stat}} - 7.2_{\mathrm{stat}})^{\circ}$ of the best-fit elliptical Gaussian, 
the fit was again performed with transposed coordinate axes of the PSF map. 
The resulting position angle is $\varphi^{*} = 47.9\deg$, corresponding to a measurable difference but still consistent with the multi-wavelength kpc jet. 
The difference $\varphi^{*} - \varphi = 4.5\deg$ is considered as a conservative estimate for the systematic error on $\varphi$. 
Because of the moderate change of the resulting position angle, we conclude that although the usage of the two-dimensional PSF in sky coordinates 
was necessary to safely present the current result, the measurement can also be conducted with a radially symmetric PSF model in this particular case. 
As an additional systematic check, we generated the PSF map for spectral indices both $\Delta \Gamma = 0.1$ softer and harder 
than the reference value of $\Gamma = 2.65$ and evaluated the influence on the extension significance and the semi-major axis width $\sigma$. 
The log-likelihood comparison of elliptical and point-like model is modified by $\Delta \mathrm{TS} = \pm 1.2$ because of this uncertainty. 
Other systematic checks include a test of the fit routine with representative toy Monte Carlo maps and the variation of the fit range. 
The fit results proved to be stable with respect to these tests, indicating that the systematic uncertainties are small compared to the statistical ones.

\begin{figure}[htbp]
\begin{center}
\includegraphics[width=0.5 \textwidth]{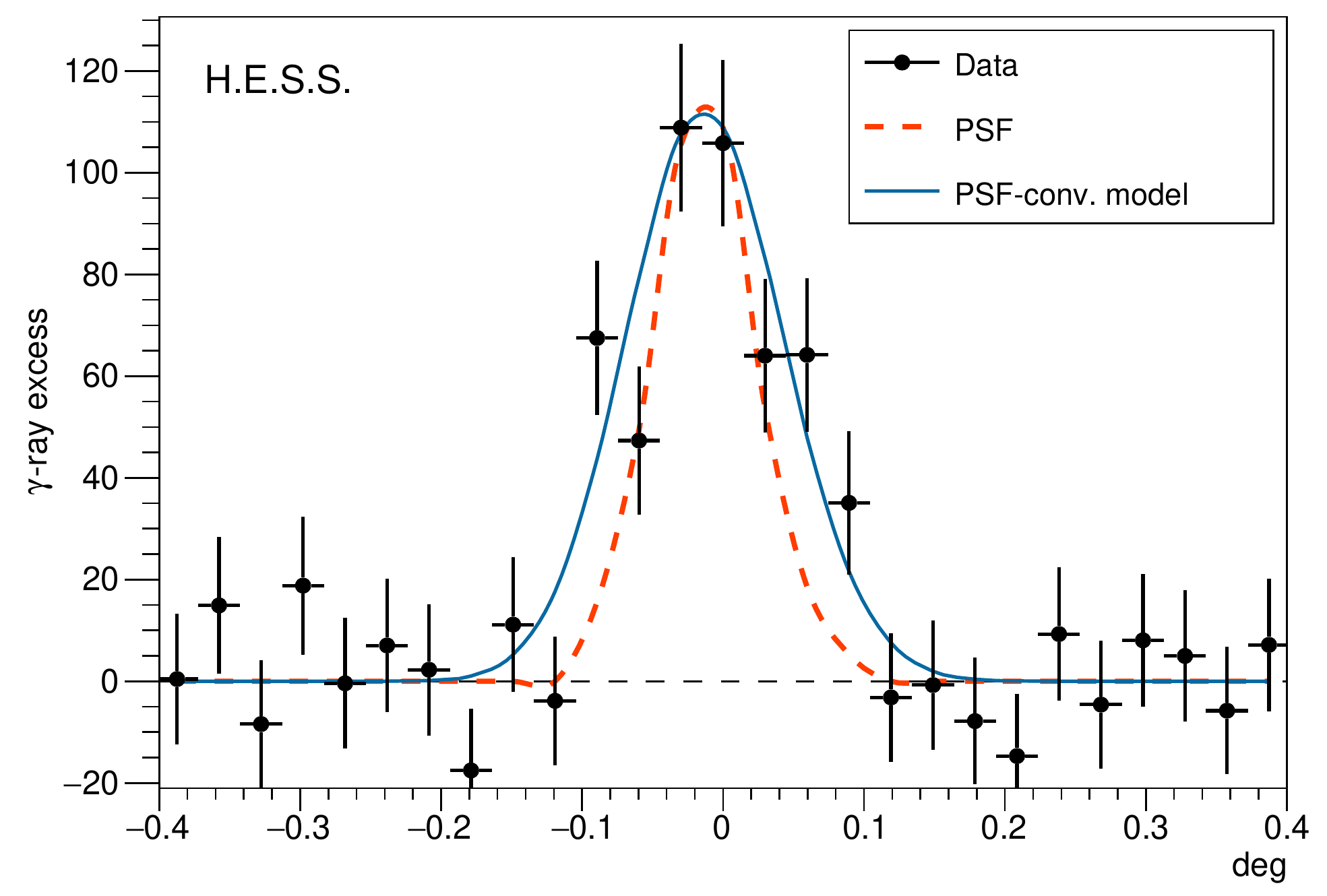}
\includegraphics[width=0.5 \textwidth]{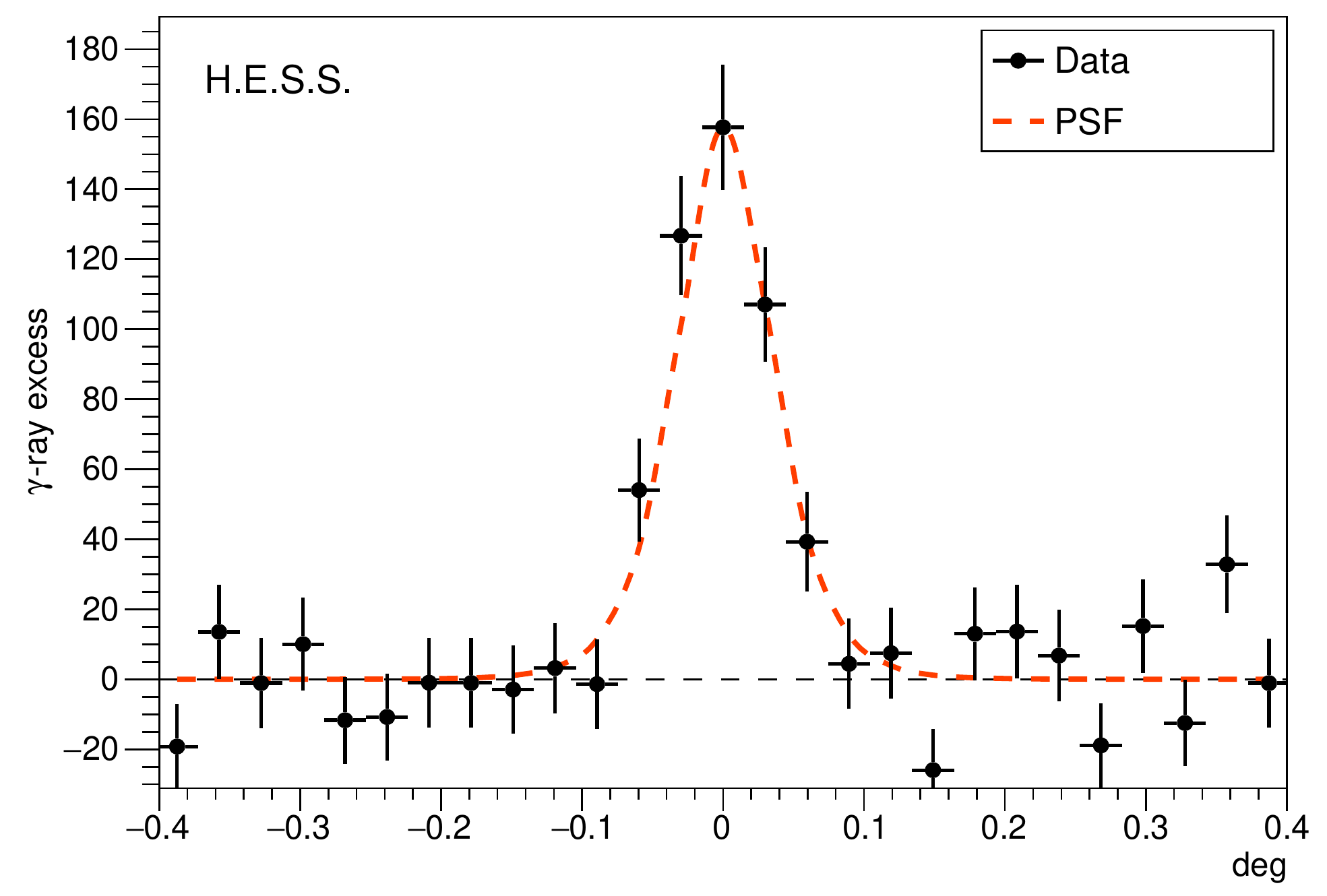}
\caption{\textbf{One-dimensional projections of VHE events.} Projections of the VHE gamma-ray emission from Centaurus A along the alignment 
of the semi-major axis obtained from the two-dimensional elliptical morphology fit (\textit{left}, 
negative values corresponding to $\varphi = 43.4\deg$ and positive ones to $\varphi + 180\deg$) 
and perpendicular to it (\textit{right}, $\varphi + 90\deg$ for negative and $\varphi + 270\deg$ for positive distances).  
The dashed red line shows the projection of the PSF on both sides. 
The blue line on the \textit{left} panel corresponds to the PSF-convolved best-fit Gaussian.
The error bars on the ordinate denote statistical uncertainties, whereas the ones along the abscissa just illustrate the bin size.}\label{projection}
\end{center}
\end{figure}

Projections of the $\gamma$-ray excess measured with H.E.S.S. along and perpendicular to the best-fit position angle of the elliptical Gaussian model are shown
in Extended Data Figure~\ref{projection}, supporting the result of the two-dimensional morphology fit. While the data are perfectly described by the PSF along the semi-minor axis (\textit{right} 
panel), the profile appears considerably extended along the semi-major axis with respect to the PSF (\textit{left} panel). 
The emission is well described by a Gaussian with the width of the semi-major axis obtained from the \textit{Sherpa} fit, convolved with the PSF. 
Additionally, the extension along this direction was determined with a $\chi^2$ fit of the Gauss-convolved PSF to the data, 
yielding $\sigma_{\mathrm{maj}} = (0.042 \pm 0.007_{\mathrm{stat}})^{\circ}$. 

All results were cross-checked with an independent calibration and analysis chain \citeMethod{Method_2014_ImPACT}, 
using the standard simulation scheme and thus a radially symmetric PSF description. 
Centaurus A is in this case detected with a slightly lower significance of $12.1\sigma$ as well as a signal to background ratio of $0.32$. 
The $68\%$ and $80\%$ containment radii of the PSF are $0.065\deg$ and $0.085\deg$, respectively. 
The best-fit parameters are consistent with those of the main analysis, but the extension significance is lowered, 
where the elliptical Gaussian model is preferred by $\mathrm{TS} = 7.7$ over the point-like one. 
This is however expected because of the higher signal to background ratio as well as the better PSF of the main analysis.

\subsection*{II. Theoretical Modelling}\label{modelling}

In the following we summarize the jet properties and the external radiation fields relevant for the SED modelling, and evaluate the 
constraints concerning acceleration and energy losses.  We then provide a reconstruction of the multi-wavelength SED along with a 
discussion of the fit parameters.\\

\noindent
{\it The large-scale Jet in Centaurus A:}
Radio observations of Centaurus A show a complex and extended morphology including a parsec-scale jet and 
counter-jet system, a one-sided kiloparsec-scale jet, and giant outer lobes whose length extends up to hundreds of 
kiloparsecs \citeMethod{Method_1983ApJ...273..128B,Method_1998A&ARv...8..237I}. Chandra X-ray observations reveal a one-sided, 
large-scale (up to $\sim4.5$ kpc in projection) jet composed both of several bright knots as well as continuous 
diffuse emission \citeMethod{Method_2000ApJ...531L...9K,Method_2002ApJ...569...54K,Method_2007ApJ...670L..81H}. In order to power the emission
from the giant outer lobes of Centaurus A, a mean kinetic jet power of the order of $L_j\sim 10^{43}-10^{44}$ 
erg/sec has been inferred \citeMethod{Method_2012A&A...542A..19Y,Method_2016A&A...595A..29S}. Constraints on the proper motion 
of jet substructures on scales of hundreds of parsecs suggests trans-relativistic jet speeds of $\sim0.5$ c 
\citeMethod{Method_2003ApJ...593..169H,Method_2019ApJ...871..248S}. Detailed analysis of the (extended) X-ray jet emission provides 
support for a stratified (fast outflow and boundary shear layer) jet model, with indications for the operation of 
a distributed acceleration mechanism in the kiloparsec-scale jet \citeMethod{Method_2006ApJ...641..158K,Method_2007ApJ...670L..81H}. 
In general, the origin of the non-thermal X-ray emission from large-scale AGN jets could be related to inverse 
Compton up-scattering of low energy photons (e.g., CMB, starlight, dust) or synchrotron emission processes, 
see ref.~\citeMethod{Method_2006ARA&A..44..463H} for review. 
 
For Centaurus A it has previously been suggested that the (knot-related) X-ray emission of its large-scale jet 
detected by the Einstein observatory likely originates from synchrotron emission, mainly based on the 
apparent spatial coincidence of the radio with the X-ray jet, and a simple comparison of the required energy 
in particles for an inverse-Compton origin of the (knot-related) emission with minimum energy estimates for 
the inner lobes \citeMethod{Method_1981ApJ...251...31F,Method_1983ApJ...273..128B}. The latter is not without assumptions, and 
the former similarities have been weakened by high-resolution Chandra observations, though the synchrotron 
interpretation is still generally favoured, e.g. refs.~\citeMethod{Method_2002ApJ...569...54K,Method_2019ApJ...871..248S}. 
Given a characteristic jet magnetic field strength of some tens of micro Gauss \citeMethod{Method_1983ApJ...273..128B}, 
a synchrotron X-ray origin would however require the presence and maintenance of ultra-relativistic electrons 
with Lorentz factors up to $\gamma\sim10^8$ \citeMethod{Method_2002ApJ...569...54K,Method_2006ApJ...641..158K}. 
As shown below, the VHE observations reported here indeed provide a clear and direct confirmation of this interpretation.\\
 
\noindent
{\it External Radiation Fields in Centaurus A:}
Optical images reveal a prominent dark band crossing the center of Centaurus A (NGC 5128). 
This dark band is related to an extended thin disk (ETD) of gas and dust that extends over several kpc and presumably originates from a merger 
event with a medium-size late-type spiral galaxy some few hundred million years ago \citeMethod{Method_2010A&A...515A..67S}. 
On smaller scales the central supermassive black hole in Centaurus A is obscured by a compact circumnuclear disk (CND) 
of size $\sim400\,\mathrm{pc} \times 200\,$pc that is somewhat warmer than the ETD \citeMethod{Method_2017A&A...599A..53I}.  
Using LABOCA measurements and archival ISO-LWS data, Wei\ss\ et al. \citeMethod{Method_2008A&A...490...77W} find that the 
SED extracted from a $80''$ aperture ($1.4\,$kpc) around the center of Centaurus A can be well described
by a two-component dust model with temperatures of $T_1=14\,$K and $T_2=30\,$K, and peak flux levels of about $80\,$Jy 
and $200\,$Jy, respectively. This amounts to a dust luminosity of the order of $L_d\sim 10^{44}\,$erg/sec. Following their findings, 
we approximate the dust emission in our model by two modified black body distributions with $F_{\nu} \propto B_{\nu}(T_{1,2})\, 
\nu^{\beta}$, where $F_{\nu}$ and $B_{\nu}(T_{1,2})$ are the flux density and the Planck function for the relevant temperature 
$T_{1,2}$, respectively. The dust emissivity index $\beta$ is taken to be 2 \citeMethod{Method_2008A&A...490...77W}. 
For the starlight contribution Abdo et al \citeMethod{Method_2010Sci...328..725A} infer a V-band luminosity for the host galaxy of Centaurus A 
of $L_V = 7.8 \times 10^{43}\,$erg/s that is similar to other estimates \citeMethod{Method_1998A&ARv...8..237I,Method_2013A&A...558A..19W}. 
The surface brightness distribution of Centaurus~A is known to closely follow an $r^{1/4}$ de Vaucouleurs' profile characteristic of 
elliptical galaxies, with effective radius $r_e=330''$ corresponding to $\simeq6\,$kpc \citeMethod{Method_1976ApJ...208..673V}. 
Feigelson et al. \citeMethod{Method_1981ApJ...251...31F} deduce an energy density in starlight photons around the X-ray jet (at the location of knot B, 
i.e. at $\sim 1\,$kpc) of the order $2\times 10^{-12}\,$erg/cm$^3$, equivalent to a local luminosity in starlight of $L_s\sim 7\times 10^{42}\,$erg/sec. 
This is comparable to an estimate based on the brightness profile and used as reference value for the modelling. 
We note that absorption of nuclear VHE emission on the starlight has also been proposed to lead to an 
isotropic kpc-scale pair halo \citeMethod{Method_stawarz2006}.\\

\begin{figure}[h!]
\begin{center}
\includegraphics[width=0.7 \textwidth]{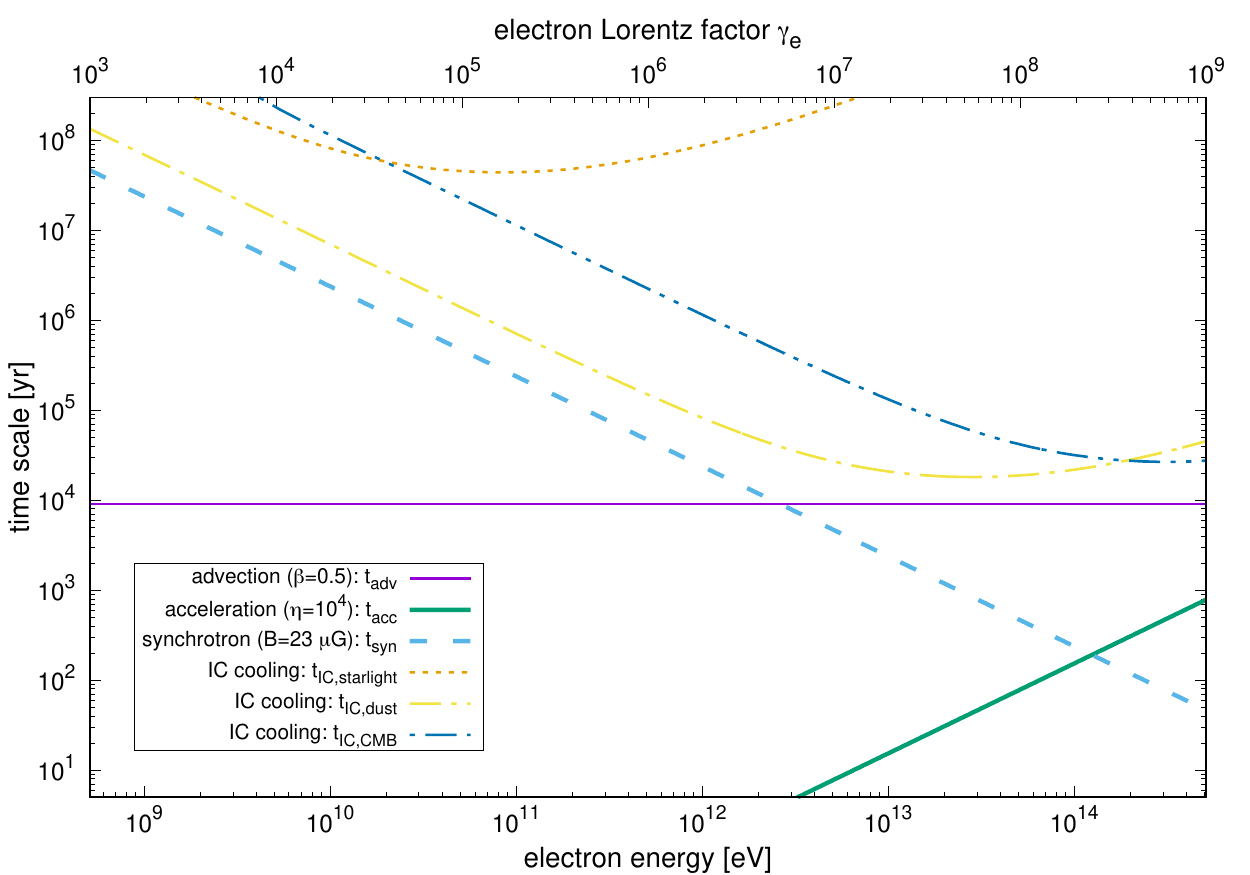}
\caption{\textbf{Relevant time scales in the jet.} Characteristic electron cooling timescales in the kpc-scale jet of Centaurus A. 
Achievable particle energies are essentially limited by synchrotron losses.}
\label{timescales}
\end{center}
\end{figure}

\noindent
 {\it Particle Energy Loss Timescales in the kpc-scale Jet:}
Energetic electrons and positrons in the large-scale jet of Centaurus A will experience synchrotron as well as IC losses on the CMB, 
starlight and dust emission. In the following we refer to electrons only as the radiation from positrons is indistinguishable from the one 
from electrons, and the positron fraction in the jet is not known. Extended Data Figure~\ref{timescales} provides an illustration of the relevant timescales 
at a location of $1.4\,$kpc, assuming a reference magnetic field of $B=23\,\mu$G, a simplified dust black body field of $30\,$K  ($L_{BB} \sim 5\times 10^{43}$ erg/sec) and 
advection with $\beta=v_j/c=0.5$ \citeMethod{Method_2003ApJ...593..169H,Method_2019ApJ...871..248S}. 
IC losses are calculated in the isotropic, mono-energetic approximation following Aharonian \&\ Atoyan~\citeMethod{Method_1981Ap&SS..79..321A}. 
Acceleration is described in terms of a fiducial acceleration efficiency $\eta$ (in units of the Larmor time). 
According to Extended Data Figure~\ref{timescales}, advection and synchrotron cooling provide the most relevant constraints. 
Synchrotron losses will become dominant above $\gamma\sim 10^6$, suggesting that the electron distribution may change its shape at around this scale. 
For the considered parameters, electron acceleration up to $\gamma_{\rm max} \sim 2.5 \times 10^8 ~(10^4/\eta)^{1/2} (23\,\mu{\mathrm{G}}/B)^{1/2}$ is possible, 
leading to a synchrotron contribution that reaches into the hard X-rays. 

Given the short synchrotron lifetimes of X-ray emitting electrons ($\leq 300\,$yr), the detection of extended X-ray emission on kpc scales 
related to synchrotron emission requires the operation of a distributed acceleration mechanism such as stochastic or 
shear particle acceleration \citeMethod{Method_2017ApJ...842...39L}. 
We note that the SED modelling results obtained below imply an electron acceleration efficiency $\eta \sim 10^4$. 
If proton acceleration was characterised by a similar efficiency, up to PeV ($10^{15}\,$eV) energies would be achievable.\\ 

\noindent
{\it SED Modelling:}
Emission from the kpc-scale jet of Centaurus A has been investigated from the radio via the infrared to the X-ray regime 
\citeMethod{Method_1981ApJ...251...31F,Method_1983ApJ...273..128B,Method_2000ApJ...531L...9K,Method_2003ApJ...593..169H,Method_2006MNRAS.368L..15H}. 
In order to study the influence of dust we focus on the inner region of the large-scale jet, i.e. on scales of $\sim (1-2)\,$kpc 
for which the dust emission profile is well known \citeMethod{Method_2008A&A...490...77W}. This demands a compromise as to the other multi-wavelength SED points. 
Hardcastle et al. \citeMethod{Method_2006MNRAS.368L..15H} have reported radio VLA ($1.4$, $4.9$ and $8.4\,$GHz), infrared Spitzer ($24$ and $5.4\,\mu$m), 
ultraviolet GALEX ($231\,$nm) and Chandra X-ray ($1\,$keV) flux measurements for three different regions outside the dust lane 
(to derive IR and UV data points) excluding compact point sources (knots). 
We use their values for the inner region (corresponding to $2.4-3.6\,$kpc in projection) to approximate the corresponding flux levels 
(green points in Fig.~\ref{SED}), noting that these values should be rather understood as upper limits. 
The high-energy ({\it Fermi}-LAT, red points) and very high energy data (H.E.S.S., blue butterfly showing the $1\sigma$-confidence band) shown in Fig.~\ref{SED} are from ref.~\citeMethod{Method_2018_CenACore}.  
These gamma-ray flux points encompass the core region ($\sim 0.1^{\circ}$) and may include a non-negligible contribution from the 
nucleus \citeMethod{Method_2018_CenACore}. This particularly holds for the sub-GeV flux points where unification models predict a significant nuclear 
(sub-parsec scale) jet contribution \citeMethod{Method_2001MNRAS.324L..33C}. 
Towards higher energies, their use as suitable reference points is justified by the absence of detected variability which would be associated with nuclear emission.\\
We consider an IC origin of the VHE emission, with the seed photons being provided by the jet emission itself 
(i.e., its low-energy synchrotron part, SSC) and the relevant external photon fields (viz. dust, CMB, starlight). 
While IC up-scattering in some of the bright knots may occur, their contribution is considered to be sub-dominant when compared with the diffuse jet 
emission.
SED modelling is done using the NAIMA package \citeMethod{Method_naima,Method_naima2010,Method_naima2014} assuming an exponential-cutoff broken power-law 
electron distribution $n(\gamma) \propto \gamma^{-\alpha_{1,2}}$ with power-law indices $\alpha_1$ and $\alpha_2$ below and above the break $\gamma_b$, 
respectively, and a super-exponential cutoff $\exp(-[\gamma/\gamma_c]^2)$ at $\gamma_c$. 
A representative model curve is given in Figure~\ref{SED}. 
The employed set of parameters is $\alpha_1=2.30$, $\alpha_2=3.85$, $\gamma_{\rm min}=100$, $\gamma_b=1.4\times 10^6$, $\gamma_c=10^8$, $B=23\,\mu$G, 
and the total energy in electrons is $W_e = 4\times 10^{53}\,$erg, see also Extended Data Table~\ref{Tab_model}. 
The results provide evidence that acceleration of electrons (and/or positrons) to ultra-relativistic energies far away from the black hole is sufficient to 
account for the observed VHE emission.

\begin{table}[h!]
\begin{center}
\begin{tabular}{lcc}
  \hline
  \hline
  \noalign{\smallskip}
  Parameter &  Notation & Value  \\
  \hline
  \noalign{\smallskip}
  Minimum Lorentz factor & $\gamma_{\rm min}$ & $10^2$  \\ 
  \noalign{\smallskip}
  Break Lorentz factor & $\gamma_b$ & $1.4\times 10^6$ \\ 
  \noalign{\smallskip}
  Cut-off Lorentz factor & $\gamma_c$ & $10^8$  \\ 
  \noalign{\smallskip}
  Power law index 1 & $\alpha_1$ & $2.30$  \\ 
  \noalign{\smallskip}
  Power law index 2 & $\alpha_2$ & $3.85$  \\ 
  \noalign{\smallskip}
  Magnetic field strength & $B$ & $23\,\mu$G  \\ 
  \noalign{\smallskip}
  Total energy in electrons & $W_e$ & $4\times 10^{53}\,$erg  \\ 
  \noalign{\smallskip}
  \hline
\end{tabular}
\caption{\textbf{Modelling parameters:} Parameters used for modelling the SED of Centaurus~A.} 
\label{Tab_model}
\end{center}
\end{table}

\begin{figure}[h!]
\begin{center}
\includegraphics[width=0.7 \textwidth]{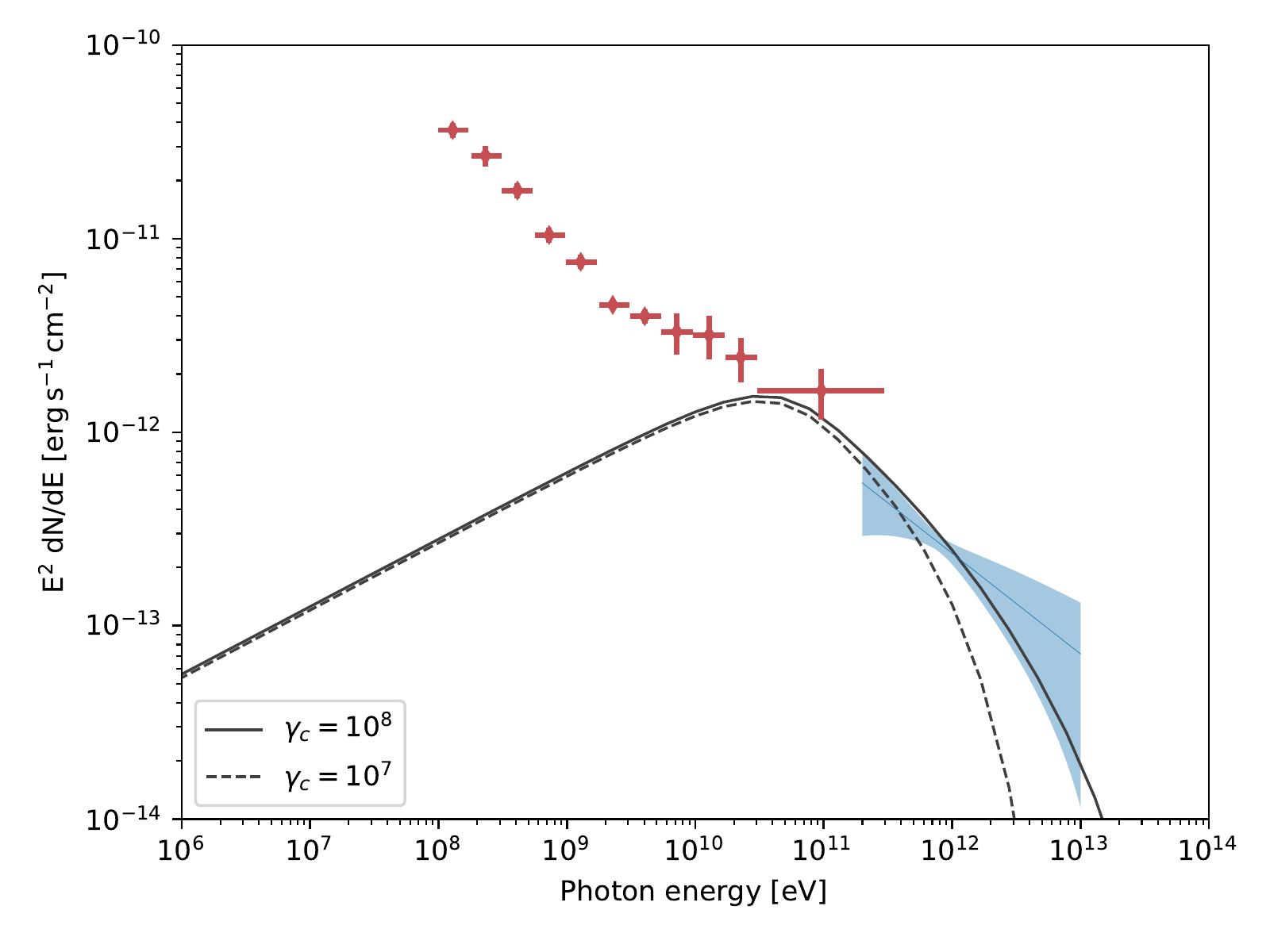} 
\caption{\textbf{Gamma-ray SED of Centaurus A.} Comparison of the resultant gamma-ray SEDs for Centaurus A including 
an earlier energy cut-off $\gamma_c m_e c^2$ for the electron distribution at $\gamma_c\simeq 10^7$ (dashed line), 
everything else being kept the same as for Figure~\ref{SED}. 
An extension of the electron distribution to $\gamma \sim 10^8$ is needed to fully account for the observed VHE spectrum. 
Red points refer to {\it Fermi}-LAT observations, usually attributed to emission from the core which is not modelled here. 
The blue-shaded butterfly represents VHE observations by H.E.S.S. 
\protect\citeMethod{Method_2018_CenACore}.}\label{SED_compare}   
\end{center}
\end{figure}

Given its large inclination and modest flow speeds $\beta=v_j/c\sim 0.5$ \citeMethod{Method_2003ApJ...593..169H,Method_2019ApJ...871..248S}, 
a Doppler factor of one has been assumed for the jet. The VHE emission in Figure~\ref{SED} is dominated by IC upscattering of dust photons, 
suggesting that strong dust emission may be a prerequisite for a large-scale FR~I jet to become bright at VHE energies \citeMethod{Method_2011MNRAS.415..133H}. 
IC scattering off a much stronger (infrared and optical) starlight photon field would not allow the reproduction of the overall spectral shape at 
VHE energies due to the reduction in efficiency for scattering occurring in the Klein-Nishina regime. 
Interestingly, our results suggest that gamma-ray emission from the large-scale jet could be the reason for the 
spectral hardening that is seen above $\sim3\,$GeV \citeMethod{Method_2018_CenACore}.
Comparison of our model curve parameters for a homogeneous zone indicates a magnetic field value that is slightly sub-equipartition; 
quoted equipartition estimates in the literature are in the range $\sim(30-60)\,\mu$G 
\citeMethod{Method_1981ApJ...251...31F,Method_1983ApJ...273..128B,Method_2002ApJ...569...54K}.\\ 
Extended Data Figure~\ref{SED_compare} shows a zoom-in version of the SED with a comparison of the resultant IC VHE
contribution of an electron distribution with an earlier cut-off, 
illustrating that electrons above $\gamma \geq 10^7$ are needed to explain the VHE emission (see Extended Data Figure~\ref{SED_compare}). 

\bibliographystyleMethod{Nature}
\bibliographyMethod{CenA_Nature_Method}

\section*{Data Availability Statement}

The raw H.E.S.S. data and the code used in this study are not public, but belong to the H.E.S.S. collaboration. All derived higher-level data that are shown in the plots will be made available on the H.E.S.S. collaboration's website upon publication of this study.

\clearpage
\section*{Full list of authors}
\par\noindent{\large

H.~Abdalla$^{\ref{NWU}}$,
R.~Adam$^{\ref{LLR}}$,
F.~Aharonian$^{\ref{MPIK},\ref{DIAS},\ref{RAU}}$,
F.~Ait~Benkhali$^{\ref{MPIK}}$,
E.O.~Ang\"uner$^{\ref{CPPM}}$,
M.~Arakawa$^{\ref{Rikkyo}}$,
C.~Arcaro$^{\ref{NWU}}$,
C.~Armand$^{\ref{LAPP}}$,
H.~Ashkar$^{\ref{IRFU}}$,
M.~Backes$^{\ref{UNAM},\ref{NWU}}$,
V.~Barbosa~Martins$^{\ref{DESY}}$,
M.~Barnard$^{\ref{NWU}}$,
Y.~Becherini$^{\ref{Linnaeus}}$,
D.~Berge$^{\ref{DESY}}$,
K.~Bernl\"ohr$^{\ref{MPIK}}$,
R.~Blackwell$^{\ref{Adelaide}}$,
M.~B\"ottcher$^{\ref{NWU}}$,
C.~Boisson$^{\ref{LUTH}}$,
J.~Bolmont$^{\ref{LPNHE}}$,
S.~Bonnefoy$^{\ref{DESY}}$,
J.~Bregeon$^{\ref{LUPM}}$,
M.~Breuhaus$^{\ref{MPIK}}$,
F.~Brun$^{\ref{IRFU}}$,
P.~Brun$^{\ref{IRFU}}$,
M.~Bryan$^{\ref{GRAPPA}}$,
M.~B\"{u}chele$^{\ref{ECAP}}$,
T.~Bulik$^{\ref{UWarsaw}}$,
T.~Bylund$^{\ref{Linnaeus}}$,
M.~Capasso$^{\ref{IAAT}}$,
S.~Caroff$^{\ref{LPNHE}}$,
A.~Carosi$^{\ref{LAPP}}$,
S.~Casanova$^{\ref{IFJPAN},\ref{MPIK}}$,
M.~Cerruti$^{\ref{LPNHE},\ref{CerrutiNowAt}}$,
T.~Chand$^{\ref{NWU}}$,
S.~Chandra$^{\ref{NWU}}$,
A.~Chen$^{\ref{WITS}}$,
S.~Colafrancesco$^{\ref{WITS}}$ \protect\footnotemark[2],
M.~Cury{\l}o$^{\ref{UWarsaw}}$,
I.D.~Davids$^{\ref{UNAM}}$,
C.~Deil$^{\ref{MPIK}}$,
J.~Devin$^{\ref{CENBG}}$,
P.~deWilt$^{\ref{Adelaide}}$,
L.~Dirson$^{\ref{HH}}$,
A.~Djannati-Ata\"i$^{\ref{APC}}$,
A.~Dmytriiev$^{\ref{LUTH}}$,
A.~Donath$^{\ref{MPIK}}$,
V.~Doroshenko$^{\ref{IAAT}}$,
L.O'C.~Drury$^{\ref{DIAS}}$,
J.~Dyks$^{\ref{NCAC}}$,
K.~Egberts$^{\ref{UP}}$,
G.~Emery$^{\ref{LPNHE}}$,
J.-P.~Ernenwein$^{\ref{CPPM}}$,
S.~Eschbach$^{\ref{ECAP}}$,
K.~Feijen$^{\ref{Adelaide}}$,
S.~Fegan$^{\ref{LLR}}$,
A.~Fiasson$^{\ref{LAPP}}$,
G.~Fontaine$^{\ref{LLR}}$,
S.~Funk$^{\ref{ECAP}}$,
M.~F\"u{\ss}ling$^{\ref{DESY}}$,
S.~Gabici$^{\ref{APC}}$,
Y.A.~Gallant$^{\ref{LUPM}}$,
F.~Gat{\'e}$^{\ref{LAPP}}$,
G.~Giavitto$^{\ref{DESY}}$,
D.~Glawion$^{\ref{LSW}}$,
J.F.~Glicenstein$^{\ref{IRFU}}$,
D.~Gottschall$^{\ref{IAAT}}$,
M.-H.~Grondin$^{\ref{CENBG}}$,
J.~Hahn$^{\ref{MPIK}}$,
M.~Haupt$^{\ref{DESY}}$,
G.~Heinzelmann$^{\ref{HH}}$,
G.~Henri$^{\ref{Grenoble}}$,
G.~Hermann$^{\ref{MPIK}}$,
J.A.~Hinton$^{\ref{MPIK}}$,
W.~Hofmann$^{\ref{MPIK}}$,
C.~Hoischen$^{\ref{UP}}$,
T.~L.~Holch$^{\ref{HUB}}$,
M.~Holler$^{\ref{LFUI}}$*,
D.~Horns$^{\ref{HH}}$,
D.~Huber$^{\ref{LFUI}}$,
H.~Iwasaki$^{\ref{Rikkyo}}$,
M.~Jamrozy$^{\ref{UJK}}$,
D.~Jankowsky$^{\ref{ECAP}}$,
F.~Jankowsky$^{\ref{LSW}}$,
A.~Jardin-Blicq$^{\ref{MPIK}}$,
I.~Jung-Richardt$^{\ref{ECAP}}$,
M.A.~Kastendieck$^{\ref{HH}}$,
K.~Katarzy{\'n}ski$^{\ref{NCUT}}$,
M.~Katsuragawa$^{\ref{KAVLI}}$,
U.~Katz$^{\ref{ECAP}}$,
D.~Khangulyan$^{\ref{Rikkyo}}$,
B.~Kh\'elifi$^{\ref{APC}}$,
J.~King$^{\ref{LSW}}$,
S.~Klepser$^{\ref{DESY}}$,
W.~Klu\'{z}niak$^{\ref{NCAC}}$,
N.~Komin$^{\ref{WITS}}$,
K.~Kosack$^{\ref{IRFU}}$,
D.~Kostunin$^{\ref{DESY}}$ ,
M.~Kraus$^{\ref{ECAP}}$,
G.~Lamanna$^{\ref{LAPP}}$,
J.~Lau$^{\ref{Adelaide}}$,
A.~Lemi\`ere$^{\ref{APC}}$,
M.~Lemoine-Goumard$^{\ref{CENBG}}$,
J.-P.~Lenain$^{\ref{LPNHE}}$,
E.~Leser$^{\ref{UP},\ref{DESY}}$,
C.~Levy$^{\ref{LPNHE}}$,
T.~Lohse$^{\ref{HUB}}$,
I.~Lypova$^{\ref{DESY}}$,
J.~Mackey$^{\ref{DIAS}}$,
J.~Majumdar$^{\ref{DESY}}$,
D.~Malyshev$^{\ref{IAAT}}$,
V.~Marandon$^{\ref{MPIK}}$,
A.~Marcowith$^{\ref{LUPM}}$,
A.~Mares$^{\ref{CENBG}}$,
C.~Mariaud$^{\ref{LLR}}$,
G.~Mart\'i-Devesa$^{\ref{LFUI}}$,
R.~Marx$^{\ref{MPIK}}$,
G.~Maurin$^{\ref{LAPP}}$,
P.J.~Meintjes$^{\ref{UFS}}$,
A.M.W.~Mitchell$^{\ref{MPIK},\ref{MitchellNowAt}}$,
R.~Moderski$^{\ref{NCAC}}$,
M.~Mohamed$^{\ref{LSW}}$,
L.~Mohrmann$^{\ref{ECAP}}$,
C.~Moore$^{\ref{Leicester}}$,
E.~Moulin$^{\ref{IRFU}}$,
J.~Muller$^{\ref{LLR}}$,
T.~Murach$^{\ref{DESY}}$,
S.~Nakashima $^{\ref{RIKKEN}}$,
M.~de~Naurois$^{\ref{LLR}}$*,
H.~Ndiyavala $^{\ref{NWU}}$,
F.~Niederwanger$^{\ref{LFUI}}$,
J.~Niemiec$^{\ref{IFJPAN}}$,
L.~Oakes$^{\ref{HUB}}$,
P.~O'Brien$^{\ref{Leicester}}$,
H.~Odaka$^{\ref{Tokyo}}$,
S.~Ohm$^{\ref{DESY}}$,
E.~de~Ona~Wilhelmi$^{\ref{DESY}}$,
M.~Ostrowski$^{\ref{UJK}}$,
I.~Oya$^{\ref{DESY}}$,
M.~Panter$^{\ref{MPIK}}$,
R.D.~Parsons$^{\ref{MPIK}}$,
C.~Perennes$^{\ref{LPNHE}}$,
P.-O.~Petrucci$^{\ref{Grenoble}}$,
B.~Peyaud$^{\ref{IRFU}}$,
Q.~Piel$^{\ref{LAPP}}$,
S.~Pita$^{\ref{APC}}$,
V.~Poireau$^{\ref{LAPP}}$,
A.~Priyana~Noel$^{\ref{UJK}}$,
D.A.~Prokhorov$^{\ref{WITS}}$,
H.~Prokoph$^{\ref{DESY}}$,
G.~P\"uhlhofer$^{\ref{IAAT}}$,
M.~Punch$^{\ref{APC},\ref{Linnaeus}}$,
A.~Quirrenbach$^{\ref{LSW}}$,
S.~Raab$^{\ref{ECAP}}$,
R.~Rauth$^{\ref{LFUI}}$,
A.~Reimer$^{\ref{LFUI}}$,
O.~Reimer$^{\ref{LFUI}}$,
Q.~Remy$^{\ref{LUPM}}$,
M.~Renaud$^{\ref{LUPM}}$,
F.~Rieger$^{\ref{MPIK}}$*,
L.~Rinchiuso$^{\ref{IRFU}}$,
C.~Romoli$^{\ref{MPIK}}$,
G.~Rowell$^{\ref{Adelaide}}$,
B.~Rudak$^{\ref{NCAC}}$,
E.~Ruiz-Velasco$^{\ref{MPIK}}$,
V.~Sahakian$^{\ref{YPI}}$,
S.~Saito$^{\ref{Rikkyo}}$,
D.A.~Sanchez$^{\ref{LAPP}}$*,
A.~Santangelo$^{\ref{IAAT}}$,
M.~Sasaki$^{\ref{ECAP}}$,
R.~Schlickeiser$^{\ref{RUB}}$,
F.~Sch\"ussler$^{\ref{IRFU}}$,
A.~Schulz$^{\ref{DESY}}$,
H.M.~Schutte$^{\ref{NWU}}$,
U.~Schwanke$^{\ref{HUB}}$,
S.~Schwemmer$^{\ref{LSW}}$,
M.~Seglar-Arroyo$^{\ref{IRFU}}$,
M.~Senniappan$^{\ref{Linnaeus}}$,
A.S.~Seyffert$^{\ref{NWU}}$,
N.~Shafi$^{\ref{WITS}}$,
K.~Shiningayamwe$^{\ref{UNAM}}$,
R.~Simoni$^{\ref{GRAPPA}}$,
A.~Sinha$^{\ref{APC}}$,
H.~Sol$^{\ref{LUTH}}$,
A.~Specovius$^{\ref{ECAP}}$,
M.~Spir-Jacob$^{\ref{APC}}$,
{\L.}~Stawarz$^{\ref{UJK}}$,
R.~Steenkamp$^{\ref{UNAM}}$,
C.~Stegmann$^{\ref{UP},\ref{DESY}}$,
C.~Steppa$^{\ref{UP}}$,
T.~Takahashi $^{\ref{KAVLI}}$,
T.~Tavernier$^{\ref{IRFU}}$,
A.M.~Taylor$^{\ref{DESY}}$*,
R.~Terrier$^{\ref{APC}}$,
D.~Tiziani$^{\ref{ECAP}}$,
M.~Tluczykont$^{\ref{HH}}$,
C.~Trichard$^{\ref{LLR}}$,
M.~Tsirou$^{\ref{LUPM}}$,
N.~Tsuji$^{\ref{Rikkyo}}$,
R.~Tuffs$^{\ref{MPIK}}$,
Y.~Uchiyama$^{\ref{Rikkyo}}$,
D.J.~van~der~Walt$^{\ref{NWU}}$,
C.~van~Eldik$^{\ref{ECAP}}$,
C.~van~Rensburg$^{\ref{NWU}}$,
B.~van~Soelen$^{\ref{UFS}}$,
G.~Vasileiadis$^{\ref{LUPM}}$,
J.~Veh$^{\ref{ECAP}}$,
C.~Venter$^{\ref{NWU}}$,
P.~Vincent$^{\ref{LPNHE}}$,
J.~Vink$^{\ref{GRAPPA}}$,
F.~Voisin$^{\ref{Adelaide}}$,
H.J.~V\"olk$^{\ref{MPIK}}$,
T.~Vuillaume$^{\ref{LAPP}}$,
Z.~Wadiasingh$^{\ref{NWU}}$,
S.J.~Wagner$^{\ref{LSW}}$,
R.~White$^{\ref{MPIK}}$,
A.~Wierzcholska$^{\ref{IFJPAN},\ref{LSW}}$,
R.~Yang$^{\ref{MPIK}}$,
H.~Yoneda$^{\ref{KAVLI}}$,
M.~Zacharias$^{\ref{NWU}}$,
R.~Zanin$^{\ref{MPIK}}$,
A.A.~Zdziarski$^{\ref{NCAC}}$,
A.~Zech$^{\ref{LUTH}}$,
A.~Ziegler$^{\ref{ECAP}}$,
J.~Zorn$^{\ref{MPIK}}$,
N.~\.Zywucka$^{\ref{NWU}}$,
}\\
{\footnotesize
* Corresponding author
\begin{enumerate}
\item Centre for Space Research, North-West University, Potchefstroom 2520, South Africa \label{NWU}  
\item Universit\"at Hamburg, Institut f\"ur Experimentalphysik, Luruper Chaussee 149, D 22761 Hamburg, Germany \label{HH}
\item Max-Planck-Institut f\"ur Kernphysik, P.O. Box 103980, D 69029 Heidelberg, Germany \label{MPIK}  
\item Dublin Institute for Advanced Studies, 31 Fitzwilliam Place, Dublin 2, Ireland \label{DIAS}  
\item High Energy Astrophysics Laboratory, RAU,  123 Hovsep Emin St  Yerevan 0051, Armenia \label{RAU} 
\item Yerevan Physics Institute, 2 Alikhanian Brothers St., 375036 Yerevan, Armenia \label{YPI} 
\item Institut f\"ur Physik, Humboldt-Universit\"at zu Berlin, Newtonstr. 15, D 12489 Berlin, Germany \label{HUB} 
\item University of Namibia, Department of Physics, Private Bag 13301, Windhoek, Namibia, 12010 \label{UNAM} 
\item GRAPPA, Anton Pannekoek Institute for Astronomy, University of Amsterdam,  Science Park 904, 1098 XH Amsterdam, The Netherlands \label{GRAPPA} 
\item Department of Physics and Electrical Engineering, Linnaeus University,  351 95 V\"axj\"o, Sweden \label{Linnaeus} 
\item Institut f\"ur Theoretische Physik, Lehrstuhl IV: Weltraum und Astrophysik, Ruhr-Universit\"at Bochum, D 44780 Bochum, Germany \label{RUB} 
\item Institut f\"ur Astro- und Teilchenphysik, Leopold-Franzens-Universit\"at Innsbruck, A-6020 Innsbruck, Austria \label{LFUI} 
\item School of Physical Sciences, University of Adelaide, Adelaide 5005, Australia \label{Adelaide} 
\item LUTH, Observatoire de Paris, PSL Research University, CNRS, Universit\'e Paris Diderot, 5 Place Jules Janssen, 92190 Meudon, France \label{LUTH} 
\item Sorbonne Universit\'e, Universit\'e Paris Diderot, Sorbonne Paris Cit\'e, CNRS/IN2P3, Laboratoire de Physique Nucl\'eaire et de Hautes Energies, LPNHE, 4 Place Jussieu, F-75252 Paris, France \label{LPNHE} 
\item Laboratoire Univers et Particules de Montpellier, Universit\'e Montpellier, CNRS/IN2P3,  CC 72, Place Eug\`ene Bataillon, F-34095 Montpellier Cedex 5, France \label{LUPM} 
\item IRFU, CEA, Universit\'e Paris-Saclay, F-91191 Gif-sur-Yvette, France \label{IRFU} 
\item Astronomical Observatory, The University of Warsaw, Al. Ujazdowskie 4, 00-478 Warsaw, Poland \label{UWarsaw} 
\item Aix Marseille Universit\'e, CNRS/IN2P3, CPPM, Marseille, France \label{CPPM} 
\item Instytut Fizyki J\c{a}drowej PAN, ul. Radzikowskiego 152, 31-342 Krak{\'o}w, Poland \label{IFJPAN} 
\item School of Physics, University of the Witwatersrand, 1 Jan Smuts Avenue, Braamfontein, Johannesburg, 2050 South Africa \label{WITS} 
\item Laboratoire d'Annecy de Physique des Particules, Univ. Grenoble Alpes, Univ. Savoie Mont Blanc, CNRS, LAPP, 74000 Annecy, France \label{LAPP} 
\item Landessternwarte, Universit\"at Heidelberg, K\"onigstuhl, D 69117 Heidelberg, Germany \label{LSW} 
\item Universit\'e Bordeaux, CNRS/IN2P3, Centre d'\'Etudes Nucl\'eaires de Bordeaux Gradignan, 33175 Gradignan, France \label{CENBG} 
\item Oskar Klein Centre, Department of Physics, Stockholm University, Albanova University Center, SE-10691 Stockholm, Sweden \label{OKC} 
\item Institut f\"ur Astronomie und Astrophysik, Universit\"at T\"ubingen, Sand 1, D 72076 T\"ubingen, Germany \label{IAAT} 
\item Laboratoire Leprince-Ringuet, \'Ecole Polytechnique, UMR 7638, CNRS/IN2P3, Institut Polytechnique de Paris, F-91128 Palaiseau, France \label{LLR} 
\item APC, AstroParticule et Cosmologie, Universit\'{e} Paris Diderot, CNRS/IN2P3, CEA/Irfu, Observatoire de Paris, Sorbonne Paris Cit\'{e}, 10, rue Alice Domon et L\'{e}onie Duquet, 75205 Paris Cedex 13, France \label{APC} 
\item Univ. Grenoble Alpes, CNRS, IPAG, F-38000 Grenoble, France \label{Grenoble} 
\item Department of Physics and Astronomy, The University of Leicester, University Road, Leicester, LE1 7RH, United Kingdom \label{Leicester} 
\item Nicolaus Copernicus Astronomical Center, Polish Academy of Sciences, ul. Bartycka 18, 00-716 Warsaw, Poland \label{NCAC} 
\item Institut f\"ur Physik und Astronomie, Universit\"at Potsdam,  Karl-Liebknecht-Strasse 24/25, D 14476 Potsdam, Germany \label{UP} 
\item Friedrich-Alexander-Universit\"at Erlangen-N\"urnberg, Erlangen Centre for Astroparticle Physics, Erwin-Rommel-Str. 1, D 91058 Erlangen, Germany \label{ECAP} 
\item DESY, D-15738 Zeuthen, Germany \label{DESY} 
\item Obserwatorium Astronomiczne, Uniwersytet Jagiello{\'n}ski, ul. Orla 171, 30-244 Krak{\'o}w, Poland \label{UJK} 
\item Centre for Astronomy, Faculty of Physics, Astronomy and Informatics, Nicolaus Copernicus University,  Grudziadzka 5, 87-100 Torun, Poland \label{NCUT} 
\item Department of Physics, University of the Free State,  PO Box 339, Bloemfontein 9300, South Africa \label{UFS} 
\item Department of Physics, Rikkyo University, 3-34-1 Nishi-Ikebukuro, Toshima-ku, Tokyo 171-8501, Japan \label{Rikkyo} 
\item Kavli Institute for the Physics and Mathematics of the Universe (WPI), The University of Tokyo Institutes for Advanced Study (UTIAS), The University of Tokyo, 5-1-5 Kashiwa-no-Ha, Kashiwa City, Chiba, 277-8583, Japan \label{KAVLI} 
\item Department of Physics, The University of Tokyo, 7-3-1 Hongo, Bunkyo-ku, Tokyo 113-0033, Japan \label{Tokyo} 
\item RIKEN, 2-1 Hirosawa, Wako, Saitama 351-0198, Japan \label{RIKKEN}
\item Now at Physik Institut, Universit\"at Z\"urich, Winterthurerstrasse 190, CH-8057 Z\"urich, Switzerland \label{MitchellNowAt} 
\item Now at Institut de Ci\`{e}ncies del Cosmos (ICC UB), Universitat de Barcelona (IEEC-UB), Mart\'{i} Franqu\`es 1, E08028 Barcelona, Spain \label{CerrutiNowAt} 
\end{enumerate}
}

\clearpage
\thispagestyle{empty}

\end{document}